\documentclass[12pt]{article}
\usepackage{amsfonts,latexsym}
\newtheorem{theorem}{Theorem}
\title{Discrete matrix Riccati equations with superposition formulas}
\author{Alexei V. Penskoi\thanks{Centre de recherches math\'ematiques, 
Universit\'e de Montr\'eal,
C.~P.~6128, succ. Centre-ville, Montr\'eal, 
Qu\'ebec, H3C 3J7, Canada {\tt e-mail: penskoi@crm.umontreal.ca}},
Pavel Winternitz\thanks{Centre de recherches math\'ematiques
et D\'epartement de math\'ematiques et de statistique, 
Universit\'e de Montr\'eal,
C.~P.~6128, succ. Centre-ville, Montr\'eal, 
Qu\'ebec, H3C 3J7, Canada {\tt e-mail: wintern@crm.umontreal.ca}}}
\date{}
\begin{document}
\maketitle

\abstract{An ordinary differential equation is said to have a
superposition formula if its general solution can be expressed as a
function of a finite number of particular solution. Nonlinear ODE's with
superposition formulas include matrix Riccati equations. Here we shall
describe discretizations of Riccati equations that preserve the
superposition formulas. The approach is general enough to include 
$q$-derivatives and standard discrete derivatives.}

\smallskip


\smallskip


\section{Introduction}
The well-known Riccati equation
\begin{equation}\label{riccati}
\dot{w}=a(t)+b(t)w+c(t)w^2
\end{equation}
arises in numerous contexts. For example, one can find particular cases 
of~(\ref{riccati}) as equations describing Darboux transformation
for the Schr\"odinger operator~\cite{ND}, as equations describing particular 
solutions of Painlev\'e equations~\cite{G}, or as B\"acklund
transformations for soliton like equations~\cite{AC,C}.

The Riccati equation has many interesting properties, one of them is the fact 
that it is a non-linear equation with a superposition 
formula. A differential equation has a superposition formula if it is 
possible to express its general solution as a function of a finite number
of  particular 
solutions and arbitrary constants. 

It is easy to prove that the Riccati equation~(\ref{riccati}) has 
a superposition formula. Let $w_0,w_1,w_2$ be three different
particular solutions of Eq.~(\ref{riccati}). 
It is well-known that using one 
particular solution one can transform~(\ref{riccati}) into a linear equation. 
Indeed, let $w=y+w_0,$ then we have $\dot{y}=(b+2cw_0)y+cy^2.$ 
Now let $y=\frac{1}{x},$ then we obtain a linear equation 
$\dot{x}=-(b+2cw_0)x-c.$ We know two particular solutions 
$x_1=\frac{1}{w_1-w_0}$ and $x_2=\frac{1}{w_2-w_0}$ of this linear equation,
hence we can find a general solution of this linear equation using $x_1$,
$x_2$ and one arbitrary constant. This gives us a formula for the general 
solution of the Riccati equation in terms of three particular
solutions $w_0,w_1,w_2$ and one arbitrary constant $c:$
\begin{equation}\label{simplesuperposition}
w=w_0+\frac{(w_2-w_0)(w_1-w_0)}{w_2-w_0+c(w_1-w_2)}.
\end{equation}

One can also consider the matrix Riccati equation
\begin{equation}\label{matrixriccati}
\dot{W}=A(t)+B(t)W+WC(t)+WD(t)W,
\end{equation}
where $W$ is a $n\times k$-matrix and $A,B,C,D$ are matrices of appropriate
sizes. This equation also has a superposition
formula~\cite{HWA,ORW}.

One can discretize the Riccati equations~(\ref{riccati}) 
and~(\ref{matrixriccati})
in different ways, but we are interested in discretizations possessing
superposition formulas. Our main result is the following. Let us consider
a two-parameter class of operators $U_{q,h}:$
\begin{equation}\label{U}
U_{q,h}f(t)=\left\{%
\begin{array}{ll}
f'(t)&\mbox{if} \quad q=1,h=0\\
\frac{f(qt+h)-f(t)}{(q-1)t+h}&\mbox{in other cases.}
\end{array}%
\right.
\end{equation}
As we shall show below the equation
\begin{equation}\label{uriccati}
U_{q,h}w(t)=a(t)+b(t)w(t)+w(qt+h)c(t)+%
w(qt+h)d(t)w(t)
\end{equation}
has a superposition formula. 
In the case where
$q=1,h=0$ we obtain the Riccati equation~(\ref{riccati}).
It can be easily seen that in the case $q=1, h\ne 0$ we obtain the 
$h$-Riccati equation
$$
\frac{w(t+h)-w(t)}{h}=a(t)+b(t)w(t)+w(t+h)c(t)+%
w(t+h)d(t)w(t),
$$
and in the case $h=0, q\ne 1$ we obtain the $q$-Riccati equation
$$
\frac{w(qt)-w(t)}{(q-1)t}=a(t)+b(t)w(t)+w(qt)c(t)+%
w(qt)d(t)w(t).
$$
Analogous results hold also in the matrix case.

It is interesting to remark that particular cases of such discretizations
appear in the same contexts as their classical analogue, for example 
as Darboux transformations for the discrete Schr\"odinger 
operator~\cite{ND,OR}, or as equations describing particular solutions
of discrete Painlev\'e equations~\cite{KK}. Thus the discretizations 
constructed in the present paper are very natural.

\section{Differential and difference equations with superposition 
formulas}\label{theory}

Let us consider a (vector) first order ordinary differential equation
\begin{equation}\label{ode}
\dot{y}^k=f^k(\mathbf{y},t),\quad k=1,\dots,n.
\end{equation}
We shall say that this equation has a superposition formula if its general 
solution $\mathbf{y}(t)$ can be expressed as a function of a finite number $m$
of particular solutions $\mathbf{y}_1,\dots,\mathbf{y}_m$ and $n$ free 
constants
\begin{equation}\label{superposition}
\mathbf{y}(t)=\mathbf{F}(\mathbf{y}_1,\dots,\mathbf{y}_m,c_1,\dots,c_n).
\end{equation}
The formula~(\ref{superposition}) is called a superposition formula
for Eq.~(\ref{ode}).

The study of ordinary differential equations with superposition formulas
has a long history that goes back to Sophus Lie. He proved the following 
fundamental theorem.
 
\begin{theorem}[Lie~\cite{L}]\label{Lietheorem}
The equation~(\ref{ode})
has a superposition formula
if and only if the function $\mathbf{f}$ has the form
$$
f^k(\mathbf{y},t)=\sum_{l=1}^rZ_l(t)\xi^k_l(\mathbf{y}),
$$
where the functions $\mathbf{\xi}_l(\mathbf{y})$ are such that the
vector fields
$$
X_l=\sum_{k=1}^n\xi^k_l(\mathbf{y})\partial_{y^k}
$$
generate a finite dimensional Lie subalgebra ${\mathfrak h}$ of
the algebra of vector fields of ${\mathbb C}^N$ or ${\mathbb R}^N,$
i.e.
$$
[X_i,X_j]=\sum_{k=1}^rC_{ijk}X_k.
$$
\end{theorem}

Difference equations with superposition formulas can be defined in 
a similar way. In order to see what can be done in the difference case,
let us reformulate the Lie theorem in different terms.

Let $G$ be a Lie group acting on a manifold $M.$ Let us consider a curve 
$g(t)$ on $G$ such that $g(t_0)=e$ and a point $u_0\in M.$ It is easy to 
find a differential equation for a curve $u(t)=g(t)\cdot u_0$ on $M.$ 
Indeed, 
$$
\dot{u}(t)=\dot{g}(t)\cdot u_0=\dot{g}(t)g(t)^{-1}g(t)\cdot u_0=%
\dot{g}(t)g(t)^{-1}\cdot u(t).
$$
Since $\dot{g}g^{-1}\in\mathfrak{g},$ we see that the following statement 
holds.

\noindent{\bf Statement 1.} A curve $u(t)=g(t)\cdot u_0,$ such that
$g(t_0)=e,$ is a solution of 
a differential equation
\begin{equation}\label{odeonM}
\dot{u}(t)=\xi(t)\cdot u(t),
\end{equation}
where $\xi=\dot{g}g^{-1}\in\mathfrak{g}.$
The point $u_0$ plays the role of an initial condition, $u(t_0)=u_0.$

Let us now consider Eq.~(\ref{odeonM}) with given $\xi(t).$
Let $u(t)$ be a solution. Let us choose a fixed value of the parameter
$t_0$ and the initial value of the solution $u_0=u(t_0).$
There exists a unique curve $g(t)\in G$ such that $\dot{g}g^{-1}=\xi$ and
$g(t_0)=e.$ We see that the curve $\tilde{u}(t)=g(t)\cdot u_0$ 
is also a solution of Eq.~(\ref{odeonM}), and it also satisfies the 
initial condition $\tilde{u}(t_0)=u_0.$ The following statement follows 
from the uniqueness theorem for solutions of ODEs.

\noindent{\bf Statement 2.} Let $\xi(t)$ be a curve on $\mathfrak{g}.$
Let $t_0$ be a fixed value of the parameter. Any solution $u(t)$ of 
equation~(\ref{odeonM}) has the form $u(t)=g(t)\cdot u_0,$ 
where $u_0\in M$ is the initial condition $u(t_0)=u_0$
and $g(t)$ is a curve on $G$ such that $g(t_0)=e.$
The curve $g(t)\in G$ is uniquely determined by the conditions 
$\dot{g}g^{-1}=\xi$ and $g(t_0)=e,$ hence $g(t)$ is the same for all
solutions $u(t)$ of equation~(\ref{odeonM}).

Statements 1 and 2 mean that Eq.~(\ref{odeonM}) has
a superposition formula. Indeed, let
$u_1(t),\dots,u_m(t)$ be particular solutions of~(\ref{odeonM}).
Let us choose an initial value of the parameter $t_0.$
It follows from Statement 2 that $u_1(t),\dots,u_m(t)$ have the form 
$u_i(t)=g(t)\cdot u_i(t_0),$
where $g(t)$ is the same for all $i.$
Then we have the following system of equations for $g(t):$
$$
\left\{%
\begin{array}{rcl}
u_1(t)&=&g(t)\cdot u_1(t_0),\\
&\dots&\\
u_m(t)&=&g(t)\cdot u_m(t_0).
\end{array}
\right.
$$
If $m$ is sufficiently large, 
$g(t)$ can be expressed in terms of $u_i(t)$ and $u_i(t_0):$
$$
g(t)=F(u_1(t),\dots,u_m(t);u_1(t_0),\dots,u_m(t_0)).
$$
It was shown by Lie that $m$ should at least
satisfy the inequality $mn\ge r,$ where $n=\dim M$ and $r$ is the same
$r$ as in Theorem~\ref{Lietheorem}, i.~e. the dimension
of the image of $\mathfrak g$ in the algebra of vector fields on $M.$

It follows that the general solution has the form
\begin{equation}\label{superpositionformula}
u(t)=g(t)\cdot u_0=F(u_1(t),\dots,u_m(t);u_1(t_0),\dots,u_m(t_0))\cdot u(t_0),
\end{equation}
the initial condition $u(t_0)$ plays the role of the arbitrary constant in the 
superposition formula~(\ref{superpositionformula}). It should be
remarked that the function $F$ in the 
superposition formula~(\ref{superpositionformula})
does not depend on the choice of $t_0,$ hence the superposition formula
is essentially the same for any choice of $t_0.$ The superposition formula
obtained by choosing another initial value of the parameter $t_1$ can be
obtained form~(\ref{superpositionformula}) by expressing the constants
$u_1(t_0),\dots,u_m(t_0),u(t_0)$ in terms of 
$u_1(t_1),\dots,u_m(t_1),u(t_1).$ 

It should also be mentioned that generally we also have some 
independence conditions on the
particular solutions in order to find $g(t).$ For example in the case
of a homogeneous linear equation of second order any solution is a linear
combination of two particular solutions under the condition
that these two solutions are linearly independent.

Statements 1 and 2 are not complicated. However, the Lie theorem also 
contains the following statement which is more difficult to prove.

\noindent{\bf Statement 3.} Every ordinary differential equation with a 
superposition formula must have the form~(\ref{odeonM}) for some group $G$
acting on some manifold $M$~\cite{L}.

A modern exposition of the proof can be found in~\cite{CGM}. 
Statements~1, 2 and 3 are exactly what Lie's theorem states. 
There is a very important thing used implicitly. 
In the proof of Statement~2 a crucial step makes use of 
the uniqueness theorem for solutions of ODEs.
We are working implicitly with some class of functions such that 
for solutions from this class this theorem holds.
In the same implicit manner we assume that equation~(\ref{odeonM}) is
an equation for which the uniqueness theorem holds. 
For differential equations the assumptions underlying the uniqueness
theorem are very natural. One can assume that
Eq.~(\ref{odeonM}) is defined by a continuous curve $\xi(t)$ and that
we are considering solutions in the class  $C^1(\mathbb{R}),$ i.~e. the class 
of differentiable functions with continuous derivative. 
For difference equations the situation is different and the
appropriate class of solutions must be carefully chosen and we need
a more rigorous definition.

\noindent{\bf Definition.} A differential (difference) equation is said to have
a superposition formula in some class of solutions if the general
solution belonging to this class can be expressed as a function
of a finite number $m$ of particular solutions 
$\mathbf{y}_1,\dots,\mathbf{y}_m,$ also belonging to this class,
and free constants
\begin{equation}\label{differencesuperposition}
\mathbf{y}=\mathbf{F}(\mathbf{y}_1,\dots,\mathbf{y}_m,c_1,\dots,c_n),
\end{equation}
where $n=\dim M.$
Eq.~(\ref{differencesuperposition}) is called a superposition formula.

The question of describing all difference equations with superposition
principles is beyond the scope of this paper. It is a very difficult question,
especially since different difference operators require the consideration of 
completely different classes of solutions. We will see examples later in 
the case of the discrete Riccati equation.
Let us concentrate on Statements 1 and 2 and see how one can
construct difference equations with superposition formulas.

We can start again from a Lie group $G,$ a manifold $M,$ a curve $g(t)\in M$ 
such that $g(t_0)=e$ and a point $u_0\in M.$ Let $u(t)=g(t)\cdot u_0.$ Let 
$U$ be a first order difference operator, for example of the form~(\ref{U}) 
defined in the Introduction. The difference derivative $Uu(t)$ can 
be defined only in a fixed coordinate system on $M.$ It is no longer 
an invariant object like the vector $\dot{u}(t).$ 
It is impossible, in general, to write something like 
$Uu(t)=Ug(t)\cdot u_0$ and 
reduce all to the Lie algebra, since there is no general chain rule for a 
difference operator and $Ug(t)$ is not, 
in general, an element of $\mathfrak{g}.$
However, in some cases we can construct 
surrogates of statements 1 and 2 in the following way. Let us assume that for 
some $G$ and $M$ (with a fixed coordinate chart) the following statements 
are true.

\noindent{\bf Statement 1'.} A curve $u(t)=g(t)\cdot u_0$ such that $g(t_0)=e$
is a solution of some difference equation
\begin{equation}\label{differenceeq}
Uu(t)=H(u(t)).
\end{equation}

\noindent{\bf Statement 2'.} Let $t_0$ be a fixed value of the parameter.
Any solution of equation~(\ref{differenceeq}), belonging to some
class of solutions, has the form $u(t)=g(t)\cdot u_0,$ where $u_0\in M$ is the
initial condition $u(t_0)=u_0$ and $g(t)$ is a curve on $G$
such that $g(t_0)=e.$ The curve $g(t)\in G$ is uniquely
determined by the condition $g(t_0)=e,$ i.~e. $g(t)$ is the same
for all solutions $u(t)$ of the equation~(\ref{differenceeq}). 

In this case we can easily prove in the same way as for ODEs
that the difference equation~(\ref{differenceeq}) has a superposition 
formula in the class of solutions mentioned in the Statement 2'. 
Moreover, this formula will be the same superposition
formula~(\ref{superpositionformula}) as 
for the differential equation, since
in both cases we are working with solutions of the same
form $u(t)=g(t)\cdot u_0.$

Let us now show how this works in the case of 
the matrix Riccati equation.

\section{Differential and difference Riccati equations}

It is easier in this Section
to consider the matrix Riccati equation from the beginning.
We will omit the word ``matrix'' for simplification.

Let us now recall how the Riccati equation arises from 
the construction described in Section~\ref{theory}. 
Consider $G=GL(N)$ and let $M$ be a Grassman manifold
$G_{n+k,k}$ of $k$-planes in an $N=(n+k)$-dimensional space. The homogeneous
coordinates of a point $p\in M$ are given by the components of an
\mbox{$(n+k)\times k$}-dimensional matrix
\begin{equation}\label{homogeneous}
\left(%
\begin{array}{c}
X\\
Y
\end{array}%
\right),
\end{equation}
where $X$ is a \mbox{$n\times k$}-matrix and 
$Y$ is a \mbox{$k\times k$}-matrix. The columns of~(\ref{homogeneous})
span a $k$-plane defining $p.$ The point $p$ is thus identified with
the equivalence class 
$\left[\left(%
\begin{array}{c}
X\\
Y
\end{array}%
\right)\right]$
under the relation
$$
\left(%
\begin{array}{c}
X\\
Y
\end{array}%
\right)\sim\left(%
\begin{array}{c}
Xh\\
Yh
\end{array}%
\right),\quad h\in GL(k),
$$
identifying different bases for the same $k$-plane. The action of an element
$$
g=\left(%
\begin{array}{cc}
M&N\\
P&Q
\end{array}%
\right)\in GL(N)
$$
upon $M$ is obtained by the projection 
$$
\pi:%
\left(%
\begin{array}{c}
X\\
Y
\end{array}%
\right)\mapsto%
\left[\left(%
\begin{array}{c}
X\\
Y
\end{array}%
\right)\right]
$$
from a linear action
$$
g\cdot%
\left(%
\begin{array}{c}
X\\
Y
\end{array}%
\right)=%
\left(%
\begin{array}{cc}
M&N\\
P&Q
\end{array}%
\right)%
\left(%
\begin{array}{c}
X\\
Y
\end{array}%
\right).
$$

On the affine subspace defined by $\det Y\ne 0$ we may define inhomogeneous
coordinates $W=XY^{-1},$ $W$ is an $n\times k$-matrix. In the inhomogeneous
coordinates the group action is given by the formula
\begin{equation}\label{inhomogeneousaction}
g\cdot W=(MW+N)(PW+Q)^{-1}.
\end{equation}

Let us now consider Eq.~(\ref{odeonM}).
In this case it reduces to Eq.~(\ref{matrixriccati}),
where $A, B, C, D$ are defined by $\xi(t)$:
\begin{equation}\label{xi}
\xi(t)=\left(%
\begin{array}{cc}
B(t)&A(t)\\
-D(t)&-C(t)
\end{array}%
\right)\in\mathfrak{gl}(N).
\end{equation}

Let us now construct a difference version of the Riccati equation.
We use the two-parameter class of operators $U_{q,h}$
defined by the formula~(\ref{U}) in the Introduction.
This class of operators can be described in an axiomatic way by the 
following properties:
\begin{enumerate}
\item $U_{q,h}\,1=0.$
\item $U_{q,h}\,(af+bg)(t)=a\,U_{q,h}\,f(t)+%
b\,U_{q,h}\,g(t),$
where $a$ and $b$ are constants.
\item $[U_{q,h},t]\,f(t)=f(qt+h).$
\end{enumerate}
For us the important property of these operators is the 
modified Leibnitz rule:
$$
U_{q,h}\,(fg)(t)=g(t)\,U_{q,h}\,f(t)+%
f(qt+h)\,U_{q,h}\,g(t).
$$

Let us proceed as described in Section~\ref{theory}. 
Let $G=GL(N)$ and $M$ be the Grassman manifold
$G_{n+k,k}$ as before, where $N=n+k.$ 
We need to choose a coordinate chart on $M:$ let us take the 
inhomogeneous coordinates described above.

Let $g(t)$ be a curve on $GL(N)$ such that $g(t_0)=e$ and $W_0$ be the
inhomogeneous coordinates of a point on $M.$
We will represent $g(t)$ as
$$
g(t)=\left(%
\begin{array}{cc}
M(t)&N(t)\\
P(t)&Q(t)
\end{array}%
\right).
$$
Let us find a difference equation for $W(t)=g(t)\cdot W_0.$ The group
action is described by the formula~(\ref{inhomogeneousaction}), so we have
$$
W(t)=(M(t)W_0+N(t))(P(t)W_0+Q(t))^{-1}.
$$
We can find $U_{q,h}\,W(t)$ directly, but it is better to proceed
in the following way. Let us choose homogeneous coordinates for $W_0,$
i.~e. matrices $X_0,$ $Y_0$ such that $X_0Y_0^{-1}=W_0.$ Then we can consider
$X(t)$ and $Y(t)$ defined by
$$
\left(%
\begin{array}{c}
X(t)\\
Y(t)
\end{array}%
\right)=g(t)\left(%
\begin{array}{c}
X_0\\
Y_0
\end{array}%
\right)=\left(%
\begin{array}{c}
M(t)X_0+N(t)Y_0\\
P(t)X_0+Q(t)Y_0
\end{array}%
\right).
$$
It is easy to see that $W(t)=X(t)Y(t)^{-1}.$ 
Since the group action in homogeneous coordinates is simply the matrix
multiplication, it is easy to see that
$$
U_{q,h}\,\left(%
\begin{array}{c}
X(t)\\
Y(t)
\end{array}%
\right)=\left(U_{q,h}\,g(t)\right)\left(%
\begin{array}{c}
X_0\\
Y_0
\end{array}%
\right)=\left(U_{q,h}\,g(t)\right)g(t)^{-1}\left(%
\begin{array}{c}
X(t)\\
Y(t)
\end{array}%
\right).
$$
Let us represent $\left(U_{q,h}\,g(t)\right)g(t)^{-1}$ 
in the following way:
$$
\left(U_{q,h}\,g(t)\right)g(t)^{-1}=\left(%
\begin{array}{cc}
B(t)&A(t)\\
-D(t)&-C(t)
\end{array}%
\right),
$$
in the case of $U_{1,0}=\frac{d}{d\,x}$ the matrix functions
$A(t),\dots,D(t)$ are the same 
as in~(\ref{xi}). We obtain the following linear difference (or differential
in the case $U_{1,0}$) equations:
\begin{eqnarray}
U_{q,h}\,X(t)&=&B(t)X(t)+A(t)Y(t),\label{1st}\\
U_{q,h}\,Y(t)&=&-D(t)X(t)-C(t)Y(t).\label{2nd}
\end{eqnarray}

It is easy to prove that for two matrices $A(t)$ and $B(t)$ we have the
identity
$$
U_{q,h}\,\left[AB^{-1}\right](t)=%
\left[U_{q,h}\,A(t)\right]B(t)^{-1}-%
\left[AB^{-1}\right](qt+h)\left[U_{q,h}\,B(t)\right]%
B(t)^{-1}.
$$
Using this identity and Eqs.~(\ref{1st},~\ref{2nd})
we can finally find a difference equation for $W(t):$
$$
U_{q,h}\,W(t)=U_{q,h}\,\left[XY^{-1}\right](t)=
$$
$$
=\left[U_{q,h}\,X(t)\right]Y(t)^{-1}-%
\left[XY^{-1}\right](qt+h)\left[U_{q,h}\,Y(t)\right]%
Y(t)^{-1}=
$$
$$
=A(t)+B(t)W(t)+W(qt+h)C(t)+W(qt+h)D(t)W(t).
$$
Thus we have found Statement 1' from section~\ref{theory} in our case:
\begin{theorem}\label{statm1'}
A function
\begin{equation}\label{solution}
W(t)=(M(t)W_0+N(t))(P(t)W_0+Q(t))^{-1},
\end{equation}
such that 
\begin{equation}\label{formofg}
g(t)=\left(%
\begin{array}{cc}
M(t)&N(t)\\
P(t)&Q(t)
\end{array}%
\right)\in GL(N)
\end{equation}
and $g(t_0)=I,$ is a solution of
the difference matrix Riccati equation
\begin{equation}\label{differencericcati}
U_{q,h}\,W(t)=%
A(t)+B(t)W(t)+W(qt+h)C(t)+W(qt+h)D(t)W(t)
\end{equation}
with the initial condition $W(t_0)=W_0.$ 

The matrices $A(t),\dots,D(t)$ are defined by the formula
\begin{equation}\label{coeff}
\left(%
\begin{array}{cc}
B(t)&A(t)\\
-D(t)&-C(t)
\end{array}%
\right)=\left[U_{q,h}\,g(t)\right]g(t)^{-1}.
\end{equation}
\end{theorem}

In the case of $U_{1,0}=\frac{d}{d\,x}$ we obtain the usual
(matrix) Riccati equation.

It is easy to see that the limit of Eq.~(\ref{differencericcati})
when $q\to1$ and $h\to0$ is the usual Riccati equation~(\ref{matrixriccati})
with the same $A(t),\dots,D(t).$ Hence we can consider 
Eq.~(\ref{differencericcati}) as a discretization of
the usual Riccati equation~(\ref{matrixriccati}).

As explained in Section~\ref{theory}, in order to prove the existence of 
a superposition formula we need to
prove Statement 2' from Section~\ref{theory} in our case, i.~e. we need
to prove that any solution of Eq.~(\ref{differencericcati})
belonging to some class of solutions
has the form~(\ref{solution}), where 
the point $W_0$ is the initial condition $W(t_0)=W_0,$
and the curve $g(t)$~(\ref{formofg})
is uniquely determined by the condition $g(t_0)=I,$ 
i.~e. $g(t)$ is the same for all solutions.

As was already explained, we need a uniqueness
theorem to prove this. This requires some assumptions on 
equation~(\ref{differencericcati}) and on the class of considered
solutions.

\section{Superposition formulas}

The theory of superposition formulas for the discrete Riccati 
equation~(\ref{differencericcati})
depends on the values of $q$ and $h.$ 
We will consider two very different cases:
the case $|q|\ne 1$ and the case $q=1, h\ne 0.$ 
We will not consider other cases since the case $q=1, h=0$ is simply
that of differential equations, and the case $|q|=1, q\ne 1$ 
would merit a separate treatment.

\subsection{Superposition formulas in the case $|q|\ne 1$}\label{qne1}

The key observation in this case is the following.
Let $T$ be the transformation $T:t\mapsto qt+h$ of the complex plane.
Then the set of points $T^n(t_0), n\in\mathbb{Z},$ has an accumulation point
$\frac{h}{1-q}.$ The well-known uniqueness 
theorem for holomorphic functions~\cite{AF} says that if a function $f,$ 
holomorphic 
in a disc, is equal to zero on a subset possessing an accumulation
point inside the disc, then $f$ is identically zero. This provides us with
a uniqueness theorem for solutions of a difference equation 
in the class of functions meromorphic in the entire 
complex plane and holomorphic in a neighborhood of
the accumulation point. Let us consider
two solutions of Eq.~(\ref{differencericcati})
from this class. Let us consider a disc such that both
solutions are holomorphic in this disc. For simplicity let us consider the case
$|q|<1.$ Let our solutions be equal in some point $t_0$ inside the disc.
Then we shall prove that two 
solutions are equal also in the points $T^n(t_0), n\in\mathbb{N}.$ 
It follows from the above mentioned uniqueness theorem
for holomorphic functions
that two solutions are identical.

On the other hand the class of functions meromorphic in the entire 
complex plane and holomorphic in a neighborhood of
the accumulation point is very natural since
if $g(t)$ is holomorphic then the solution $W(t)$~(\ref{solution}) is,
in general, a meromorphic function.

This leads us to the following theorem.

\begin{theorem}\label{th3}
Let $|q|\ne 1.$
The discrete Riccati equation~(\ref{differencericcati}) with holomorphic
coefficients $A(t), B(t), C(t), D(t)$ has a superposition formula in the class
of functions meromorphic in the entire complex plane and holomorphic 
in a neighborhood of the point $\frac{h}{1-q}.$
\end{theorem}

\noindent{\bf Proof.} Let us remark that the change of coordinates
$t=t'+\frac{h}{1-q}$ permits us to reduce 
Eq.~(\ref{differencericcati}) to an equation of the same form 
but with $h=0.$ Hence we shall consider only this case, the case 
of $q$-derivative. Let us also remark that the change of coordinates
$t=\frac{1}{q}t'$ permits us to reduce further to an equation of the same form 
but with $|q|<1.$

Let us consider the formula~(\ref{coeff}) as 
a difference equation for the $N\times N$-matrix $g(t).$
We can rewrite it as
\begin{equation}\label{eqforg}
U_{q,0}\,g(t)=%
\left(%
\begin{array}{cc}
B(t)&A(t)\\
-D(t)&-C(t)
\end{array}%
\right)g(t).
\end{equation}
It is a homogeneous linear $q$-difference equation for $g(t).$ 
Let us also fix some $t_0$ and consider the initial problem $g(t_0)=I.$

It can be proven by methods similar to those used in Ref.~\cite{A}
that there exists
a unique solution $g(t)$ holomorphic in the entire complex plane. Let us
prove that $g(t)$ is non-degenerate, i.~e. $g(t)\in GL(N).$ Suppose that
in some point $t_1$ we have $\det g(t_1)=0.$ We can rewrite 
equation~(\ref{eqforg}) as
$$
g(qt)=((q-1)t%
\left(%
\begin{array}{cc}
B(t)&A(t)\\
-D(t)&-C(t)
\end{array}%
\right)%
+I)g(t).
$$
We see that if $g(t)$ is degenerate then $g(qt)$ is degenerate. It follows
that for the holomorphic function $\det g(t)$ we have 
$\det g(q^nt_1)=0, n\in\mathbb{N}.$ This implies $g(t)\equiv 0,$ but
it contradicts the initial condition $g(t_0)=I.$ Hence $g(t)\in GL(N).$

Let $W_1(t)$ be a solution of Eq.~(\ref{differencericcati})
meromorphic in the entire complex plane and holomorphic in 
a neighborhood of the point $0.$

Let us choose some $R>0$ such that the following conditions hold
in the disc $|t|<R:$
\begin{enumerate}
\item $W_1(t)$ is holomorphic; 
\item The matrix 
$$
I-(q-1)t(C(t)+D(t)W_1(t))
$$
is non-degenerate.
\end{enumerate}
Such $R$ exists since $C(t), D(t)$ and $W_1(t)$ are holomorphic in zero.

Let us choose some $t_0$ such that $|t_0|<R.$ Let $W_0=W_1(t_0)$ and let
$$
g(t)=\left(%
\begin{array}{cc}
M(t)&N(t)\\
P(t)&Q(t)
\end{array}%
\right)\in GL(N)
$$
be the solution of~(\ref{eqforg}) with the initial condition
$g(t_0)=I.$ Let us consider the function 
\begin{equation}\label{gsol}
W_2(t)=g(t)\cdot W_0=(M(t)W_0+N(t))(P(t)W_0+Q(t))^{-1}.
\end{equation}
It follows from Theorem~\ref{statm1'}
that the function~(\ref{gsol}) satisfies 
Eq.~(\ref{differencericcati})
and the initial condition $W_2(t_0)=W_0.$
Let us note that $W_2(t)$ is meromorphic in the entire complex plane.

We can rewrite Eq.~(\ref{differencericcati}) in the form
$$
W(qt)[I-(q-1)t(C(t)+D(t)W(t))]=(q-1)t(A(t)+B(t)W(t))+W(t).
$$

Since the matrix $I-(q-1)t(C(t)+D(t)W(t))$ is invertible in the disc
$|t|<R$ and $|q|<1,$ we have
$$
W_1(qt_0)=[(q-1)t(A(t_0)+B(t_0)W_0)+W_0][I-(q-1)t(C(t_0)+D(t_0)W_0)]^{-1}.
$$
The same is true for $W_2(qt),$ so $W_2(qt_0)=W_1(qt_0).$ 
We can prove in analogous way that $W_2(q^nt)=W_1(q^nt),n\in\mathbb{N}.$ 

Since $W_1(t)$ is holomorphic in a neighborhood of $0$, we have
$$
\lim_{n\to\infty}W_2(q^nt_0)=\lim_{n\to\infty}W_1(q^nt_0)=%
W_1(0)
$$
and this implies that $W_2(t)$ cannot have a pole in $0.$
Since both $W_1(t)$ and $W_2(t)$ are holomorphic in some neighborhood 
of $0$ and their values are the same in
the sequence of points having $0$ as an accumulation point, 
we obtain $W_1(t)\equiv W_2(t).$ Thus any solution of
Eq.~(\ref{differencericcati}) 
meromorphic in the entire complex plane and holomorphic in 
a neighborhood of zero
has the form~(\ref{gsol}).

This is ``nearly'' Statement 2' from 
Section~\ref{theory}. ``Nearly'' because the initial point $t_0$ cannot
be arbitrary, but should satisfy the condition $|t_0|<R$ and this
$R$ depends on the solution.
However it permits us to find a superposition formula. Indeed, let
$W_1(t),\dots,W_m(t)$ be particular solutions
meromorphic in the entire complex plane and holomorphic in 
a neighborhood of zero.
For each $W_i$ we can find its own constant $R_i$ and take 
$R=\min(R_1,\dots,R_m).$

Let us choose an initial parameter value $t_0$ such that $|t_0|<R.$
It follows that $W_1(t),\dots,W_m(t)$ have the form 
$W_i(t)=g(t)\cdot W_i(t_0),$
where $g(t)$ is the same for all $i.$
As in Section~\ref{theory} we have the system of equations 
for $g(t).$
If $m$ is sufficiently large and 
$W_i(t)$ satisfy some additional independence conditions necessary for
$g(t)$ to be expressed in terms of $W_i(t)$
and $W_i(t_0),$ we have
$$
g(t)=F(u_1(t),\dots,u_m(t);u_1(t_0),\dots,u_m(t_0)).
$$
It follows that the general solution holomorphic in the disc 
$|t|<|t_0|+\varepsilon$ has the form
\begin{equation}\label{qsuperpositionformula}
u(t)=g(t)\cdot u_0=F(u_1(t),\dots,u_m(t);u_1(t_0),\dots,u_m(t_0))\cdot u(t_0),
\end{equation}
the initial condition $u(t_0)$ plays a role of the arbitrary constant in the 
superposition formula~(\ref{qsuperpositionformula}). 
Since the function $F$
does not depend on the choice of $t_0,$ we can consider a general 
solution holomorphic in a disc of radius less then $R.$ If we 
decrease $R$ and repeat our construction of the superposition formula,
the resulting superposition formula will be the same.
This finishes the proof. $\Box$

As we already mentioned in Section~\ref{theory}, the resulting
superposition formula is the same as in the case of usual
Riccati equation~(\ref{matrixriccati}). These superposition formulas
can be found in~\cite{HWA,ORW}.

\subsection{Superposition formulas in the case $q=1,$ $h\ne0$}

By rescaling $t=ht'$ we can transform Eq.~(\ref{differencericcati})
into the same equation with $q=1, h=1.$

This case is very different from the case $|q|\ne 1.$ It is difficult
to find a natural class of solutions defined in the entire 
complex plane. Since
the set $\{t_0+nh,n\in\mathbb{N}\}$ has no accumulation point, 
the uniqueness theorem for holomorphic functions cannot be applied. 
In this situation we have, for example, a uniqueness theorem for entire 
functions of growth order $1$ and of normal type,
but it is very unnatural. We already mentioned in the beginning
of Subsection~\ref{qne1} that solutions of the Riccati equations are in
general meromorphic. But the situation becomes very natural if we consider
functions defined only on $\mathbb{Z},$ i.~e. in integer points.

\begin{theorem} The discrete Riccati equation
\begin{equation}\label{hriccati}
W(n+1)-W(n)=A(n)+B(n)W(n)+W(n+1)C(n)+W(n+1)D(n)W(n)
\end{equation}
such that the matrices
\begin{equation}\label{matrix}
\left(%
\begin{array}{cc}
I+B(n)& A(n)\\
-D(n)& I-C(n)
\end{array}%
\right)
\end{equation}
are non-degenerate for all $n$ has a superposition formula in the class
of solutions defined on all $\mathbb{Z}.$
\end{theorem}

\noindent{\bf Proof.} Let us consider the formula~(\ref{coeff}) as 
a difference equation for the $N\times N$-matrix $g(t).$ 
We can rewrite it as
$$
g(n+1)=
\left(%
\begin{array}{cc}
I+B(n)& A(n)\\
-D(n)& I-C(n)
\end{array}%
\right)g(n).
$$
Let us also fix some $n_0$ and consider the initial problem $g(n_0)=I.$
Since the matrix~(\ref{matrix}) is non-degenerate, 
$$
g(n)=\left(%
\begin{array}{cc}
M(n)&N(n)\\
P(n)&Q(n)
\end{array}%
\right)
$$
is uniquely defined
for all $n$ and $g(n)\in GL(N).$

Let us consider a solution $W_1(n)$ defined on all $\mathbb{Z}.$
Let us use the same $n_0$ and let $W_0$ be the initial condition 
$W_0=W_1(n_0).$
We can consider a function
\begin{equation}\label{hsol}
W_2(n)=g(n)\cdot W_0=(M(n)W_0+N(n))(P(n)W_0+Q(n))^{-1}.
\end{equation}
It is also a solution of Eq.~(\ref{hriccati}) with the initial condition
$W_2(n_0)=W_0.$

Eq.~(\ref{hriccati}) can be rewritten in the
following forms:
\begin{eqnarray*}
W(n+1)\left[I-C(n)-D(n)W(n)\right]&=&A(n)+B(n)W(n)+W(n),\\
\left[I+B(n)+W(n+1)D(n)\right]W(n)&=&W(n+1)-A(n)-W(n+1)C(n).
\end{eqnarray*}
It follows that if for all $n$ we have
\begin{equation}\label{hcond}
\det[I-C(n)-D(n)W_1(n)]\ne0,\quad\det[I+B(n-1)+W_1(n)D(n-1)]\ne 0,
\end{equation}
then for all $n$ we can find $W_1(n)$ and $W_2(n)$  
starting from the initial condition $W_0=W_1(n_0)=W_2(n_0)$ and
we can see that $W_1(n)\equiv W_2(n).$ We see that all the solutions satisfying
the conditions~(\ref{hcond}) are of the form~(\ref{hsol}). This implies
in the same way as in the proof of
Theorem~\ref{th3} that equation~(\ref{hriccati})
has a superposition formula in the class of solutions defined on all
$\mathbb{Z}$ and satisfying~(\ref{hcond}). Moreover,
a general solution satisfies conditions~(\ref{hcond}) since they 
represent only a countable number of inequalities (labeled by $n$).
$\Box.$

In the case of the differential
Riccati equations we have a superposition formula in the class of
$C^1(\mathbb{R})$-solutions. However, for some particular choice of the 
constants this
formula gives us solutions outside this class, for example with poles. 
The same situation arises in the case of equation~(\ref{hriccati}). We
have a superposition formula in the class of solutions defined on all 
$\mathbb{Z},$ but for some particular choices of
the constants this formula gives us 
solutions which are singular in some points. 

For example, let us consider
$$
g(t)=\left(%
\begin{array}{cc}
1+t^2&-t^2\\
t^2&1-t^2
\end{array}%
\right)\in GL(2).
$$
This curve gives us solutions
$$
w(t)=\frac{(u_0-1)t^2+u_0}{(u_0-1)t^2+1}
$$
of the equation
$$
\dot{w}(t)=-2t+4tw(t)-2tw(t)^2
$$
with initial condition $w(0)=u_0.$
For $u_0\ge1$ these solutions belong to $C^1(\mathbb{R})$ and we can write a 
superposition formula. If we take three particular solutions
$w_0,w_1,w_2$ corresponding to $u_0=1,u_0=2$ and $u_0=3,$
respectively, we obtain 
from~(\ref{simplesuperposition}) the formula for the general solution belonging
to $C^1(\mathbb{R}):$
$$
w(t)=\frac{2t^2+4-c}{2t^2+2-c}.
$$
If, for example, $c=4,$ we obtain a solution
$$
w(t)=\frac{t^2}{t^2-1},
$$
which is singular at $t=\pm1.$

In the difference case the same curve gives us solutions
$$
w(n)=\frac{(u_0-1)n^2+u_0}{(u_0-1)n^2+1}
$$
of the equation
$$
w(n+1)-w(n)=-2n-1+(2n+1)w(n)+(2n+1)w(n+1)-(2n+1)w(n+1)w(n).
$$
We obtain the same superposition formula
$$
w(n)=\frac{2n^2+4-c}{2n^2+2-c}
$$
valid for the solutions defined on all $\mathbb{Z}.$ It also gives
us solutions with singularities. For example for $c=4$ we have
$$
w(n)=\frac{n^2}{n^2-1},
$$
which is singular at $n=\pm1.$
We see that the superposition formula 
can be used to avoid singularities in numerical computations,
as already pointed out in Ref.~\cite{RW}.

\section{Conclusions}
We have constructed very natural discrete matrix
Riccati equations with superposition formulas. This was done for a family
of difference operators general enough to include both $q$-derivatives and 
standard discrete derivatives. 

The key idea was considering superposition
formulas in an appropriate class of solutions.  Which class depends on 
the type of difference operator used. It turns out that in the case of the
$q$-derivative it is natural to consider solutions meromorphic in the
entire complex plane and holomorphic in a neighborhood of zero, and
in the case of the standard discrete derivative, it is natural to consider
solutions defined only in integer points.

The discretization of Riccati equations of this article should be compared
with a previously proposed discretization of linearizable
equations, including the Riccati one~\cite{GRW,TW}. There the curve in a Lie
algebra, i.~e. $\xi(t)$ of Eq.~(\ref{odeonM}), was replaced by a curve 
in the corresponding Lie group $G$ and this gave a linearizable mapping.
E.~g. for the Riccati equation~(\ref{riccati}), the corresponding
discretization would be
\begin{equation}\label{other}
w(t+h)=[g_{11}(t)w(t)+g_{12}(t)][g_{21}(t)w(t)+g_{22}(t)]^{-1}.
\end{equation}
Here $h$ is, on one hand, a lattice spacing, on the other hand a group
parameter. Thus, the whole class of Riccati equations is replaced by the
above class of homographic mappings. If equation~(\ref{other})
has constant coefficients, then the relation between the Riccati mapping
and the Riccati equation is simple:
$$
g=\left(%
\begin{array}{cc}
g_{11}&g_{12}\\
g_{21}&g_{22}
\end{array}%
\right)=e^{Lh},\quad L\in sl(2),\quad g\in SL(2),
$$
i.~e.: $g_{11}=1+hb+\dots,$ $g_{12}=ha+\dots,$ $g_{21}=-hd+\dots,$
$g_{22}=1-hc+\dots,$ and for $h\to0$ (\ref{other}) reduces to
$$
\dot{w}(t)=a+(b+c)w(t)+dw^2(t).
$$
For variable coefficients the relation between a specific Riccati mapping
and a specific Riccati equation is difficult to establish (i.~e. it is 
necessary to integrate the Riccati equation explicitly). For all details
we refer to the original articles~\cite{GRW,TW}. We also mention
that discrete integrable mappings of the type~(\ref{other}) have
been used to discretize the Pinney equation~\cite{RSW}.

The relation between the discrete and continuous equations of this article, on 
the other hand, is quite straightforward. Given the ODE~(\ref{matrixriccati}),
we give the discrete equation~(\ref{differencericcati}) explicitly, involving 
the same functions $A(t),\dots,D(t).$ Thus 
equation~(\ref{differencericcati}) can serve as the basis for numerical 
approximations for differential equations.
This would complement a previous use of superposition formulas
in numerical analysis~\cite{RW}.

We would like to comment on the relation between the discrete Riccati
equations and the Runge-Kutta method. Statement 2 of Section~\ref{theory}
states that in the class of $C^1(\mathbb{R})$-solutions 
the differential Riccati equation~(\ref{matrixriccati}) is equivalent 
to the linear matrix equation
\begin{equation}\label{eqforgdiff}
\dot{g}(t)=%
\left(%
\begin{array}{cc}
B(t)&A(t)\\
-D(t)&-C(t)
\end{array}%
\right)g(t)
\end{equation}
with the initial condition $g(t_0)=I.$
Applying the Runge-Kutta method to Eq.~(\ref{eqforgdiff}), we obtain
the difference equation
$$
U_{1,h}g(t)=%
\left(%
\begin{array}{cc}
B(t)&A(t)\\
-D(t)&-C(t)
\end{array}%
\right)g(t)
$$
which is in a similar way equivalent to the difference Riccati equation
defined by the operator $U_{1,h}.$ As it was explained before this will
also give us singular solutions.

Equivalently we can say that all $C^1(\mathbb{R})$-solutions of the Riccati 
equation~(\ref{matrixriccati}) have the form $W(t)=X(t)Y(t)^{-1},$
where $X(t)$ and $Y(t)$ are solutions of the linear system
\begin{equation}\label{XY}
\frac{d}{dt}\left(%
\begin{array}{c}
X(t)\\
Y(t)
\end{array}%
\right)=\left(%
\begin{array}{cc}
B(t)&A(t)\\
-D(t)&-C(t)
\end{array}%
\right)\left(%
\begin{array}{c}
X(t)\\
Y(t)
\end{array}%
\right).
\end{equation}
Applying the Runge-Kutta method to Eq.~(\ref{XY})
gives us the equation
\begin{equation}\label{hXY}
U_{1,h}\,\left(%
\begin{array}{c}
X(t)\\
Y(t)
\end{array}%
\right)=\left(%
\begin{array}{cc}
B(t)&A(t)\\
-D(t)&-C(t)
\end{array}%
\right)\left(%
\begin{array}{c}
X(t)\\
Y(t)
\end{array}%
\right).
\end{equation}
Introducing the inhomogeneous coordinates (the matrix elements
of $W=XY^{-1}$) we transform~(\ref{hXY}) into the discrete
Riccati equation~(\ref{differencericcati}) with $q=1,h\ne0.$

In other words, the $h$-discretization of the matrix
Riccati equation, presented in this article, not only
preserves the superposition formula. It is also equivalent
to applying the Runge-Kutta method to the associated linear 
system~(\ref{XY}).

\section*{Acknowledgments}

A.~P. is grateful to the Centre de recherches
math\'ematiques, Universit\'e de Montr\'eal, for its hospitality.
The research of P.~W. was partly supported by research grants from NSERC of
Canada, and FQRNT du Qu\'ebec.

\end{document}